# Personal Data Processing for Behavioural Targeting: Which Legal Basis?

Dr. Frederik Zuiderveen Borgesius

Frederikzb[at]cs.ru.nl

**Abstract:**

- The European Union Charter of Fundamental Rights only allows personal data processing if a data controller has a legal basis for the processing.

- This paper argues that in most circumstances the only available legal basis for the processing of personal data for behavioural targeting is the data subject's unambiguous consent.

- Furthermore, the paper argues that the cookie consent requirement from the e-Privacy Directive does not provide a legal basis for the processing of personal data.

- Therefore: even if companies could use an opt-out system to comply with the e-Privacy Directive's consent requirement for using a tracking cookie, they would generally have to obtain the data subject's unambiguous consent if they process personal data for behavioural targeting.

[1] The paper is based on, and includes sentences from, the PhD research of the author [F. J. Zuiderveen Borgesius, Improving Privacy Protection in the Area of Behavioural Targeting (Kluwer Law International 2015)]. The author thanks Kristina Irion, Stefan Kulk, Nico van Eijk, Joris van Hoboken, and the anonymous reviewer for their helpful comments and suggestions.

**Table of Contents**

# INTRODUCTION

In Europe, discussions about the legal requirements for behavioural targeting, a type of online marketing, often focus on the e-Privacy Directive's consent requirement for tracking cookies and similar technologies.[2] But in many cases, the right to protection of personal data also applies, namely when behavioural targeting entails the processing of personal data. The European Union Charter of Fundamental Rights only allows personal data processing if the data controller has a legal basis for the processing, such as consent.[3] The central question for this paper is which legal basis data controllers can rely on to process personal data for behavioural targeting, and which requirements follow from that conclusion. That question has received little attention in the literature.[4]

---

[2] Directive 2002/ 58/EC of the European Parliament and of the Council of 12 July 2002 concerning the processing of personal data and the protection of privacy in the electronic communications sector (Directive on privacy and electronic communications), last amended by Directive 2009/136/EC of the European Parliament and of the Council of 25 November 2009 (O.J. L 337 11) (hereinafter 'e-Privacy Directive'). Unless otherwise noted, this paper refers to the consolidated 2009 version of the e-Privacy Directive.

[3] Charter of Fundamental Rights of the European Union (OJ C 364 of 18 December 2000).

First this paper gives a brief introduction to behavioural targeting and the relevance of data protection law for this marketing technique. The legal basis for personal data processing, as required by the Charter and the Data Protection Directive, is discussed next.[5] I argue that companies usually cannot base personal data processing for behavioural targeting on the legal basis necessity for performance of a contract, or on the legal basis necessity for the controller's legitimate interests. Therefore companies must generally obtain the data subject's unambiguous consent for personal data processing for behavioural targeting.

I argue that article 5(3) of the e-Privacy Directive does not provide a legal basis for the processing of personal data. Article 5(3) of the e-Privacy Directive (sometimes called the cookie provision[6]) requires consent for storing or accessing information on a device of a user or subscriber. Some companies suggest that they can use an opt-out system to comply with article 5(3) of the e-Privacy Directive. However, even if companies could obtain consent for cookies that way, this paper shows that companies would generally be required to obtain the data subject's unambiguous consent if they process personal data for behavioural targeting.

The paper focuses on the rules in the Data Protection Directive and the e-Privacy Directive, rather than on their implementation into national law. Also outside the paper's scope are questions regarding trans-border data flows and the territorial scope of European rules.[7] Whether informed consent is a good privacy protection measure is questionable, but that question falls outside this paper's scope.[8]

---

[4] Papers discussing the legal basis for personal data processing for behavioural targeting include: J Koëter, 'Behavioral Targeting en Privacy: een Juridische Verkenning van Internetgedragsmarketing' [Behavioural Targeting and Privacy: a Legal Exploration of Behavioural Internet Marketing] (2009) Tijdschrift voor Internetrecht 2009-4; P Traung, 'EU Law on Spyware, Web Bugs, Cookies, etc. Revisited: Article 5 of the Directive on Privacy and Electronic Communications' (2010) 31 Business Law Review 2010-31; B Van Der Sloot and FJ Zuiderveen Borgesius, 'Google and Personal Data Protection' in A. Lopez-Tarruella (ed), *Google and the Law* (T.M.C. Asser Press/Springer 2012).
[5] Directive 95/46/EC of the European Parliament and of the Council of 24 October 1995 on the Protection of Individuals with Regard to the Processing of Personal Data and on the Free Movement of Such Data (O.J. L 281 31) (hereinafter 'Data Protection Directive').
[6] See for instance: P Hustinx, The Relationship between the proposed Data Protection Regulation and the e-Privacy Directive (Speech at GSMA-ETNO seminar 'The Data Protection Reform - The Consumer Perspective') Brussels, 16 October 2012.

## BEHAVIOURAL TARGETING

In a common arrangement for online advertising, advertisers only pay website publishers if somebody clicks on an ad. Click-through rates are low: "less than 1 click per 1000 impressions."[9] In other words, when an ad is shown to a thousand people, on average less than one person clicks on it. Behavioural targeting, a type of electronic marketing, was developed to increase the click-through rate on ads. Behavioural targeting involves monitoring people's online behaviour, and using the collected information to show people individually targeted advertisements. Massive amounts of data are collected about hundreds of millions of people for behavioural targeting.

In a simplified example of behavioural targeting, three parties are involved: an Internet user, a website publisher, and an advertising network. Advertising networks are companies that serve ads on thousands of websites, and can recognise people when they browse the web. If somebody often visits websites about electronic gadgets, and ad network might infer that the person is a technology enthusiast. The ad network might display advertising for electronic gadgets when that person visits a website. When visiting that same website at the same time, somebody who is profiled as a travel enthusiast might see ads for hotels.

Ad networks often use tracking cookies, small text files that are stored on a user's computer to recognise that computer. With a tracking cookie, an ad network can follow an Internet user across all websites on which it serves ads. Through almost every popular website tracking cookies are stored; through some websites dozens of them.[10] Behavioural targeting companies use many other tracking technologies as well, such as flash cookies and device fingerprinting.[11]

---

[7] See on the territorial scope of EU rules regarding behavioural targeting: F Debusseré, 'The EU E-Privacy Directive: A Monstrous Attempt to Starve the Cookie Monster' (2005) International Journal of Law and Information Technology 2005-13 70.
[8] See on that topic for instance: P Blume, 'The Inherent Contradictions in Data Protection Law' (2012) International Data Privacy Law 2(1) 26; S Barocas S and H Nissenbaum, 'Big Data's End Run around Anonymity and Consent', in J Lane et al (eds), *Privacy, Big Data, and the Public Good: Frameworks for Engagement* (Cambridge University Press 2014).
[9] See e.g. D Chaffey, 'Display advertising clickthrough rates' 21 April 2015 <http://www.smartinsights.com/internet-advertising/internet-advertising-analytics/display-advertising-clickthrough-rates/> accessed 25 May 2015.

*Internet of Things*

Currently behavioural targeting happens mostly when people use a computer or a smart phone. But the borders between offline and online are becoming fuzzier.[12] Phrases such as ubiquitous computing, the Internet of Things, and ambient intelligence have been used to describe – or promote – that development.[13] If objects are connected to the Internet, data collected through those objects could be used for behavioural targeting. An article in the Persuasive Computing journal calls "ubiquitous advertising (…) the killer application for the 21st Century".[14]

To illustrate: an Italian company sells mannequins with built-in cameras. The company's website says that the mannequins "reveal important details about your customers: age range; gender; race; number of people and time spent."[15] A drinks machine in Japan uses a camera to estimate age and gender of the user, to recommend drinks.[16] In the UK, some billboards with facial recognition technology adapt their images to the people looking at the billboard.[17] In the future, perhaps we might see behavioural targeting in the physical space, like in the movie *Minority Report*.[18] Google has said: "a few years from now, we and other companies could be serving ads and other content on refrigerators, car dashboards, thermostats, glasses, and watches, to name just a few possibilities."[19]

---

[10] CJ Hoofnagle and N Good, 'The Web Privacy Census' (October 2012) <http://law.berkeley.edu/privacycensus.htm> accessed 13 May 2015.
[11] CJ Hoofnagle et al, 'Behavioral Advertising: The Offer You Cannot Refuse' (2012) 6(2) Harvard Law & Policy Review 273.
[12] M Hildebrandt, 'The Rule of Law in Cyberspace' (Translation of Inaugural Lecture at Radboud University Nijmegen, 22 December 2011) <http://works.bepress.com/mireille_hildebrandt/48/> accessed 13 May 2015, p. 11. See also A Daly, 'The Law and Ethics of 'Self Quantified' Health Information: An Australian Perspective' (2015) International Data Privacy Law 5(2) 144.
[13] M Weiser, 'Ubiquitous Computing' (Hot Topics) (1993), IEEE Computer 26(10) 71; A Bassi et al, 'Internet of Things Strategic Research Roadmap', in O Vermesan and P Friess (eds), *Internet of Things Global Technological and Societal Trends* (River 2011); Philips Research, 'What Is Ambient Intelligence?' (2014) <www.research.philips.com/technologies/projects/ami/> accessed 13 May 2015.
[14] J Krumm, 'Ubiquitous Advertising: The Killer Application for the 21st Century' (2011) IEEE Pervasive Computing 10(1) 66.
[15] Almax, 'EyeSee Mannequin' (2012) <www.almax-italy.com/en-US/ProgettiSpeciali/EyeSeeMannequin.aspx> accessed 5 January 2014.
[16] E Lies, 'Japan Vending machine Recommends Drinks to Buyers' (14 November 2010) <www.reuters.com/article/2010/11/15/us-japan-machines-idUSTRE6AE0G720101115> accessed 13 May 2015.
[17] Y Chen, 'Facial Recognition Billboard Only Lets Women See the Full Ad' (21 February 2012) <www.psfk.com/2012/02/facial-recognition-billboard.html#!weiJ9> accessed 13 May 2015.

## THE RELEVANCE OF DATA PROTECTION LAW FOR BEHAVIOURAL TARGETING

Data protection law is relevant for behavioural targeting as far as behavioural targeting entails the processing of personal data, information about an identified or identifiable person.[20] The Article 29 Working Party, an advisory body in which national Data Protection Authorities cooperate, says behavioural targeting generally entails personal data processing, even if a company cannot tie a name to the data it holds about an individual. If a company uses data to "single out" an individual, or to distinguish an individual within a group, these data are personal data, according to the Working Party.[21] That view has met both criticism and approval.[22] The opinions of the Working Party are not legally binding, but they are influential. Judges and national Data Protection Authorities often follow the Working Party's interpretation.[23] This paper concerns situations in which behavioural targeting entails the processing of personal data.

---

[18] S Spielberg, 'Minority Report' (2002) <www.imdb.com/title/tt0181689/> accessed 13 May 2015.

[19] Google, Letter to United States Securities and Exchange Commission, 20 December 2013, <www.sec.gov/Archives/edgar/data/1288776/000128877613000074/filename1.htm> accessed 13 May 2015.

[20] Article 1(1) and 2(a) of the Data Protection Directive.

[21] See e.g. Article 29 Working Party, 'Opinion 2/2010 on online behavioural advertising' (WP 171), 22 June 2010, p. 9.

[22] Criticism: GJ Zwenne, De verwaterde privacywet [Diluted Privacy Law], Inaugural Lecture of Prof. Dr. GJ Zwenne to the Office of Professor of Law and the Information Society at the University of Leiden on Friday, 12 April 2013 (Leiden University 2013); approval: P De Hert and S Gutwirth, 'Regulating Profiling in a Democratic Constitutional State', in M Hildebrandt and S Gutwirth (eds), *Profiling the European Citizen* (Springer 2008).

[23] See generally on the Working party: S Gutwirth and Y Poullet, 'The contribution of the Article 29 Working Party to the construction of a harmonised European data protection system: an illustration of 'reflexive governance'?', in VP Asinari and P Palazzi (eds), *Défis du Droit à la Protection de la Vie Privée. Challenges of Privacy and Data Protection Law* (Bruylant 2008).

The Working Party says ad networks and website publishers are often joint controllers, as they jointly determine the purposes and means of the processing.[24] The paper only discusses situations in which a company that uses behavioural targeting is the data controller.[25] For ease of reading, this paper also refers to 'companies', instead of data controllers, and to 'people' and 'persons' instead of 'data subjects'.

The Charter of Fundamental Rights of the European Union only allows personal data processing if the controller has a legal basis for the processing.[26] The legal bases are listed in article 7 of the Data Protection Directive.[27] For the private sector the most relevant legal bases are: necessity for the performance of a contract, necessity for the controller's legitimate interests, and the data subject's unambiguous consent. Article 7 of the Data Protection Directive only concerns the legal basis for personal data processing – a controller that has a legal basis for processing must still comply with the directive's other provisions.[28] In the following we will see that in most cases, unambiguous consent is the only available legal basis for personal data processing for targeted marketing.

## NECESSITY FOR THE PERFORMANCE OF A CONTRACT

A first legal basis that a company can rely on for personal data processing is article 7(b) of the Data Protection Directive: necessity for the performance of a contract. It is sometimes suggested that Internet users "pay" for websites and other Internet services with their personal data.[29] As the Interactive Advertising Bureau US puts it, "visiting a web site is a commercial act, during which a value exchange occurs. Consumers receive content, and in exchange are delivered [targeted] advertising."[30] From an economic perspective (rather than from a legal perspective), consenting to behavioural targeting can be seen as entering into a market transaction with a company.[31]

---

[24] Article 29 Working Party 2010, 'Opinion 2/2010 on online behavioural advertising' (WP 171), 22 June 2010, p. 11.
[25] The data controller determines the purposes and means of the personal data processing (article 2(d), and is responsible for compliance (article 6(2)(b) and 23(1) of the Data Protection Directive). The processor processes personal data on behalf of the controller (article 1(e)).
[26] Article 8(2) of the Charter of Fundamental Rights of the European Union.
[27] Article 6 of the proposal for a Data Protection Regulation copies the same legal bases without major revisions (European Commission, Proposal for a Regulation of the European Parliament and of the Council on the protection of individuals with regard to the processing of personal data and on the free movement of such data (General Data Protection Regulation), 2012/0011 (COD)).
[28] CJEU, Case C-468/10 and C-469/10 ASNEF [2011], ECLI:EU:C:2011:777, par. 71.
[29] See for a critique on such claims: KJ Strandburg, 'Free Fall: the Online Market's Consumer Preference Disconnect' (2013) University of Chicago Legal Forum 2013, 95.

But from a legal perspective, an indication of wishes, or an "intention to be legally bound", of both parties is required to enter a contract.[32] In general contract law, mere silence does not constitute an indication of wishes. According to the Vienna Sales Convention for instance, "[a] statement made by or other conduct of the offeree indicating assent to an offer is an acceptance. Silence or inactivity does not in itself amount to acceptance".[33] Several proposals for European contract law use the same phrase.[34] Indeed, the results would be strange if the law did allow a seller to interpret a customer's silence as an indication of wishes. A car dealer could demand payment if a customer did not object to an offer to buy a car.

Many companies cannot base personal data processing for behavioural targeting on a contract. For instance, if an ad network collects information about people without them being aware, the ad network has not entered a contract with those people, as they have neither received nor accepted an offer to enter a contract.[35]

However, in some circumstances a company could try to argue that it can base personal data processing for behavioural targeting on the legal basis necessity for performing a contract. For example, a company that provides a social network site could argue that the user entered a contract when opening an account, and that behavioural targeting "is necessary for the performance of a contract" with the user.[36] The "contract" would imply that the user discloses personal data, in exchange for using the social network site. Indeed, European social network providers have suggested that behavioural targeting is necessary for a contract with their users:

> Whether analyzing user data for ad targeting or suggesting individual services is lawful is a controversial topic. We would highly appreciate if the future legal framework for processing of user data would clarify that these ways of analyzing and using user data do not necessarily require consent but rather are part of the processing that is necessary for the performance of a contract to which the data subject is party.[37]

---

[30] R Rothenberg, 'Has Mozilla Lost Its Values?' (Interactive Advertising Bureau US) (16 July 2013) <www.iab.net/iablog/2013/07/has-mozilla-lost-its-values.html> accessed 13 May 2015.
[31] See: A Acquisti, 'The Economics of Personal Data and the Economics of Privacy' (Background Paper for the Conference: The Economics of Personal Data and Privacy: 30 Years after the OECD Privacy Guidelines) (2010) <www.oecd.org/Internet/ieconomy/46968784.pdf> accessed 13 May 2015.
[32] J Smits, 'The Law of Contract' in J Hage and B Akkermans (eds), *Introduction to Law* (Springer 2014), p. 56.
[33] Article 18(1) of the Vienna Convention on International Sale of Goods.
[34] The same phrase is used in article II 4:204(2) of the Draft Common Frame of Reference (Principles, Definitions and Model Rules of European Private Law), and article 34 (of Annex 1) of the Proposal for a regulation of the European Parliament and of the Council on a Common European Sales Law (COM(2011) 635 final).
[35] Article 29 Working Party, Letter to Google (signed by 27 national Data Protection Authorities), 16 October 2012 <www.cnil.fr/fileadmin/documents/en/20121016-letter_google-article_29-FINAL.pdf> Appendix: <www.cnil.fr/fileadmin/documents/en/GOOGLE_PRIVACY_POLICY-_RECOMMENDATIONS-FINAL-EN.pdf> accessed 13 May 2015; College bescherming persoonsgegevens (Dutch DPA), 'Investigation into the combining of personal data by Google, Report of Definitive Findings' (z2013-00194) (November 2013) <https://cbpweb.nl/sites/default/files/downloads/mijn_privacy/en_rap_2013-google-privacypolicy.pdf> accessed 13 May 2015, p. 85.

Apart from the question whether the data subject has entered a contract, article 7(b) of the Data Protection Directive requires that the processing is "necessary" for performing that contract. The European Court of Justice says in the *Huber* case that "the concept of necessity (…) has its own independent meaning in Community law."[38] The European Court of Human Rights, says "[t]he adjective 'necessary' is not synonymous with 'indispensable', neither has it the flexibility of such expressions as 'admissible', 'ordinary', 'useful', 'reasonable' or 'desirable' (…)."[39]

---

[36] In some cases, the user of a social network site could be seen as a data controller, but I will leave this complication aside for this paper (see Article 29 Working Party, 'Opinion 5/2009 on online social networking' (WP 163) 12 June 2009).
[37] European Social Networks, 'Response to the commission ´s public consultation on the comprehensive approach on personal data protection in the European Union' (14 January 2011) <http://ec.europa.eu/justice/news/consulting_public/0006/contributions/organisations/europeansocialnetworks_en.pdf> accessed 13 May 2015, p. 5.
[38] ECJ, Case C-524/06 Huber [2008] ECLI:EU:C:2008:724, par. 52.
[39] Silver and Others v United Kingdom App no 5947/72; 6205/73; 7052/75; 7061/75; 7107/75; 7113/75; 7136/75 (ECHR, 25 March 1983), par 97.

Some authors suggest that "necessary" in data protection law should be interpreted the same way as "necessary" in the case law of the European Court of Human Rights.[40] But caution is needed when interpreting the case law from Strasbourg and Luxembourg. In *Huber*, the state was the data controller. The state did not invoke the legal basis necessity for performance of a contract (article 7(b)), but another legal basis: necessity for a task in the public interest (article 7(e)).[41] It could be argued that people's fundamental rights primarily need protection against the state, rather than against companies. In that view, companies should have more leeway than the state. This interpretation would suggest that "necessary" must be interpreted more leniently when applying article 7(b) (contract), than when applying article 7(e) (public interest). On the other hand, the state should aim to work for the common good, while companies aim for profit. This would suggest that a company should have less leeway.[42]

Without taking sides in this debate: it is clearly not enough if a behavioural targeting company finds it helpful or profitable to process personal data – the concept of necessity requires more. As Kuner notes, Data Protection Authorities "are not likely to view the processing of data as 'necessary' to perform the contract unless such processing is truly central and unavoidable in order to complete the transaction."[43] Until recently, the Working Party had been silent on the appropriate legal basis for personal data processing for behavioural targeting. In 2014 the Working Party said that necessity for performance of a contract is not an appropriate legal basis for behavioural targeting.[44] In sum, it seems implausible that companies can base processing for behavioural targeting on the legal basis necessity for performance of a contract.

The analysis becomes more complicated if a company uses the same personal data for providing its service and for behavioural targeting. Suppose a company offers a smartphone app with a personalised news service. The app analyses the user's reading habits to recommend news articles. Processing some personal data (the user's reading habits tied to a unique identifier) is necessary for performing the contract, as the app can only offer its personalised news service by analysing those data. But for providing the personalised news service it is not necessary to use the same personal data for targeted advertising. Hence, if the company wants to use the same data to target ads to the user, it needs a separate legal basis for the processing of personal data for behavioural targeting.[45]

---

[40] See e.g. LA Bygrave and DW Schartum, 'Consent, Proportionality and Collective Power' in Gutwirth S et al (eds), *Reinventing Data Protection?* (Springer 2009), p. 163.
[41] But see Bygrave, who suggests "necessary" in other data protection law provisions should probably be interpreted the same (LA Bygrave, *Data Privacy Law. An International Perspective* (Oxford University Press 2014), p. 150).
[42] See S Gutwirth, *Privacy and the Information Age* (Rowman & Littlefield 2002), p. 38.
[43] C Kuner, *European Data Protection Law: Corporate Regulation and Compliance* (Oxford University Press 2007), p. 234-235 (internal footnote omitted).
[44] Article 29 Working Party, 'Opinion 06/2014 on the notion of legitimate interests of the data controller under article 7 of Directive 95/46/EC' (WP 217) 9 April 2014, p. 17.

Similar reasoning applies to some "Internet of Things" scenarios. Say a company offers a fridge that registers when the consumer runs out of milk. The fridge automatically orders new milk, which is delivered to the consumer's house. Processing some personal data, such as the consumer's address, is necessary to deliver the milk. Processing those personal data can be based on the legal basis necessity for performance of a contract. But if the fridge company wants to use the same milk consumption data to target ads to the consumer, the company needs a separate legal basis for the processing for targeted marketing.

One difference between the legal bases consent (article 7(a)) and necessity for performance of a contract (article 7(b)) is that the procedural requirements for consent in data protection law are stricter than the procedural requirements for many contracts. For example, in general contract law, terms and conditions are often part of the contract.[46] But as discussed below, the Working party says that companies cannot obtain consent for personal data processing through a privacy statement or terms and conditions.[47]

In conclusion, it is unlikely that companies can rely on the legal basis necessity for performance of a contract for the processing of personal data for behavioural targeting.

---

[45] See Article 29 Working Party, 'Opinion 02/2013 on apps on smart devices' (WP 202) 27 February 2013, p. 13.
[46] See: K Zweigert and H Kötz, *An Introduction to Comparative Law* (Tony Weir tr, 3rd edn, OUP 1998), p. 331.
[47] See in this paper the section "Data subject's consent to personal data processing". See also Article 29 Working Party, 'Opinion 15/2011 on the definition of consent' (WP 187) 13 July 2011, p. 33-34.

## LEGITIMATE INTERESTS OF THE CONTROLLER

A second legal basis that a company can invoke for personal data processing is necessity for the legitimate interests of the controller (article 7(f)), also called the balancing provision. In brief, a controller can rely on this provision when personal data processing is necessary for the legitimate interests of the controller, or of a third party to whom the data are disclosed, unless those interests are overridden by the data subject's interests or fundamental rights.[48]

Can companies base personal data processing for behavioural targeting on the legitimate interests of the controller? Let's take a simple example: an ad network tracks people's browsing behaviour over thousands of websites, to compile individual profiles, to target ads to individuals.

A preliminary question is whether the ad network has a legitimate interest.[49] The ad network could invoke its "freedom to conduct a business in accordance with Union law and national laws and practices", as protected by the Charter of Fundamental Rights of the European Union.[50] The Advocate General of the European Court of Justice confirms that online marketing relates to the freedom to conduct a business.[51] But this freedom is not absolute; it must be balanced against other fundamental rights, such as privacy and data protection rights. A company that breaches data protection law or other laws cannot successfully invoke its right to conduct a business; its business would not be "in accordance with Union law and national laws".

The legitimate interests provision mentions "the legitimate interests pursued by the controller *or by the third party or parties to whom the data are disclosed*".[52] If an ad network allows advertisers to target ads to specific individuals (identified through cookies for instance), it essentially rents out access to those individuals. From a data protection law perspective, such a practice should be seen as a type of data disclosure. The processing definition speaks of "disclosure by transmission, dissemination *or otherwise making available*."[53] The ad network makes data available for advertisers, including when it does not provide them with a copy of the data. Korff notes that list rental is a type of data disclosure, and by analogy his conclusion can be applied to ad networks.[54] In any case, the analysis of the legitimate interests provision remains roughly the same, regardless of whether a company invokes its own interests, or those of third parties. Let's assume that the ad network in our example has a legitimate interest.

---

[48] See generally: P Balboni and others, 'Legitimate interest of the data controller New data protection paradigm: legitimacy grounded on appropriate protection' (2013) International Data Privacy Law, 3(4) 244.
[49] See Article 29 Working Party 2014, 'Opinion 06/2014 on the notion of legitimate interests of the data controller under article 7 of Directive 95/46/EC' (WP 217) 9 April 2014, p. 24-29.
[50] Article 16 of the EU Charter of Fundamental Rights.
[51] Opinion AG (ECLI:EU:C:2013:424) for CJEU, case C-131/12, Google Spain [2014] ECLI:EU:C:2014:317, par. 95. See also Article 29 Working Party, 'Opinion 06/2014 on the notion of legitimate interests of the data controller under article 7 of Directive 95/46/EC' (WP 217) 9 April 2014, p. 25: marketing is a legitimate interest.
[52] Article 7(f) of the Data Protection Directive. That directive defines 'third party" in article 2(f).

Having a legitimate interest is not enough for a company to rely on the legitimate interests provision – the processing must be "necessary." The question of necessity can be divided in two steps, subsidiarity and proportionality. Regarding subsidiarity: it seems questionable whether tracking people's browsing behaviour is the least intrusive manner for the ad network to enable advertisers to promote their products. For instance, contextual advertising is possible without tracking people's behaviour. An example of contextual advertising is displaying ads for law books on websites about law. But an ad network that specialises in behavioural targeting could try to argue that tracking people is necessary for its business.

Furthermore, Data Protection Authorities might see large-scale tracking for targeted advertising as disproportionate. After analysing how Data Protection Authorities apply the proportionality principle, Kuner concludes that "the risk of legal problems caused by application of the proportionality principle can be particularly high [in cases such as] the large-scale collection of data over the Internet."[55]

---

[53] Article 2(b) of the Data Protection Directive.
[54] D Korff D, *Data Protection Laws in the European Union* (Federation of European Direct Marketing and Direct Marketing Association 2005), p. 63. With list rental, a list broker sends leaflets to a set of people, but the advertiser does not receive a copy of the list.
[55] C Kuner, 'Proportionality in European Data Protection Law and its Importance for Data Processing by Companies' (2008) 7(44) BNA Privacy & Security Law Report, p. 1620 (capitalisation adapted).

In principle behavioural targeting would be possible without large-scale data collection. Experimental behavioural targeting systems exist that don't involve sharing one's browsing behaviour with a company. For example, a browser plug-in called Adnostic builds a profile based on the user's browsing behaviour, and uses that profile to target ads. Minimal information leaves the user's device, as the behavioural targeting happens in the user's browser.[56] Mozilla is experimenting with a similar system for the Firefox browser.[57] As behavioural targeting would be possible without large-scale data collection, it could be seen as disproportionate if companies collect large amounts of personal data for behavioural targeting. However, scholars and regulators rarely make that argument.[58] In sum, the necessity test is not a trivial hurdle to take. But let's assume the ad network in our example passes this hurdle.

The next question is whether the data subject's fundamental rights or interests override the company's interests.[59] When balancing the interests of the company and the data subject, it must be taken into account that privacy and data protection rights are fundamental rights.[60] It follows from European Court of Human Rights case law that people have a reasonable expectation of privacy regarding their Internet use.[61] And a Council of Europe resolution says that online tracking is a privacy threat.[62] In addition, surveys show that many people find tracking for behavioural targeting intrusive.[63]

---

[56] S Barocas et al, 'Adnostic: Privacy Preserving Targeted Advertising' (2010) NDSS. See also: <http://crypto.stanford.edu/adnostic/> accessed 13 May 2015.

[57] J Scott, 'A User Personalization Proposal for Firefox' (Mozilla Labs Updates from the edge of the Web) (25 July 2013) <https://blog.mozilla.org/labs/2013/07/a-user-personalization-proposal-for-firefox/> accessed 13 May 2015.

[58] Acquisti, an economist, makes an argument along those lines (A Acquisti, 'The Economics of Personal Data and the Economics of Privacy' (Background Paper for the Conference: The Economics of Personal Data and Privacy: 30 Years after the OECD Privacy Guidelines) (2010) <www.oecd.org/Internet/ieconomy/46968784.pdf> accessed 13 May 2015, p. 42-43.

[59] CJEU, Case C-468/10 and C-469/10 ASNEF [2011], ECLI:EU:C:2011:777, par 38.

[60] CJEU, Case C-468/10 and C-469/10 ASNEF [2011], ECLI:EU:C:2011:777, par. 41. See also ECJ, Case C-465/00, C-138/01 and C-139/01 Österreichischer Rundfunk [2003], ECLI:EU:C:2003:294, par. 68; CJEU, Case C-131/12 Google Spain [2014] ECLI:EU:C:2014:317, par. 74.

[61] *Copland v United Kingdom* App no 62617/00 (ECHR 3 April 2007), par. 42.

[62] Parliamentary Assembly, Resolution 1843, The protection of privacy and personal data on the Internet and online media, 7 October 2011, par 18.6.

[63] See: J Turow et al, 'Americans Reject Tailored Advertising and Three Activities that Enable it' (29 September 2009) <http://ssrn.com/abstract=1478214> accessed 13 May 2015. In Europe, seven out of ten people are concerned that companies might use data for new purposes such as targeted advertising without informing them (European Commission, 'Special Eurobarometer 359: Attitudes on data

But the data subject's rights aren't absolute. The European Court of Justice calls for "a fair balance (…) between the various fundamental rights and freedoms protected by the EU legal order."[64] When balancing the opposing interests, "the seriousness of the infringement of the data subject's fundamental rights" can be taken into account.[65] The Working party says that other factors to consider include the sensitivity of the data, the scale of data collection, the reasonable expectations of the data subject, and the risks involved.[66] For instance, mobile location data are rather sensitive.

Companies can never rely on necessity for the legitimate interests of the controller (article 7(f) of the Data Protection Directive) as a legal basis for processing special categories of data, such as data regarding political opinions or health. Unless a specified exception applies, processing special categories of data is prohibited, or, depending on the national implementation law, only allowed after the data subject's explicit consent.[67]

A few authors have already concluded that companies cannot rely on necessity for the legitimate interests of the controller (article 7(f) of the Data Protection Directive) as a legal basis for behavioural targeting that involves tracking people over multiple websites.[68] Our behavioural targeting example discussed above concerns a simple form of behavioural targeting, which involves tracking people's browsing behaviour. Some behavioural targeting companies go further, for instance by merging different data sets to enrich user profiles. If a practice is more invasive, there is less chance that a company can rely on the legitimate interests provision (article 7(f)). In sum, the most convincing

---

protection and electronic identity in the European Union' (2011) <http://ec.europa.eu/public_opinion/archives/ebs/ebs_359_en.pdf> accessed 13 May 2015, p 146.
[64] CJEU, Case C-468/10 and C-469/10 ASNEF [2011], ECLI:EU:C:2011:777, par. 43.
[65] CJEU, Case C-468/10 and C-469/10 ASNEF [2011], ECLI:EU:C:2011:777, par. 44.
[66] See Article 29 Working Party, 'Opinion 06/2014 on the notion of legitimate interests of the data controller under article 7 of Directive 95/46/EC' (WP 217) 9 April 2014, p. 33-43.
[67] Article 8 of the Data Protection Directive.
[68] See P Traung P, 'EU Law on Spyware, Web Bugs, Cookies, etc. Revisited: Article 5 of the Directive on Privacy and Electronic Communications' (2010) 31 Business Law Review 216, p. 218; L Moerel, 'Big Data protection. How to make the draft EU Regulation on Data Protection future proof' (inaugural lecture) (14 February 2014) <www.debrauw.com/wp-content/uploads/NEWS%20-%20PUBLICATIONS/Moerel_oratie.pdf> accessed 13 May 2015, p. 58. The Dutch government comes to the same conclusion. See for an English translation of the relevant remarks of the Dutch legislator: College bescherming persoonsgegevens (Dutch DPA), 'Investigation into the combining of personal data by Google, Report of Definitive Findings' (z2013-00194) (November 2013) <https://cbpweb.nl/sites/default/files/downloads/mijn_privacy/en_rap_2013-google-privacypolicy.pdf> accessed 13 May 2015, p. 81, footnote 294.

view is that personal data processing for behavioural targeting that involves tracking people over various Internet services cannot be based on article 7(f) of the Data Protection Directive, necessity for the legitimate interests of the controller.

In exceptional circumstances, companies might be allowed to base personal data processing for behavioural targeting on the legal basis necessity for the legitimate interests of the controller (article 7(f) of the Data Protection Directive). For instance, perhaps an online bookstore that tracks people's behaviour within its website to provide recommendations could rely on the legitimate interests provision. Arguably people are more likely to understand what happens when they see behaviourally targeted ads that are based on browsing behaviour within one website, than when they are confronted with targeted ads based on tracking over various Internet services.[69]

The UK Information Commissioner's Office (ICO) said in 2010 that behavioural targeting generally entails the processing of personal data. About the legal basis for processing, the ICO added: "there are alternatives to consent".[70] Presumably, the ICO hinted at necessity for the controller's legitimate interests (article 7(f) of the Data Protection Directive) as a possible legal basis. However, in 2013 the Working Party said that the data subject's unambiguous consent is the only appropriate legal basis for behavioural targeting.[71]

Behavioural targeting is a type of direct marketing, as confirmed in a code of conduct of the Federation of European Direct and Interactive Marketing: "Direct marketing in the on-line environment refers to one-to-one marketing activities where individuals are targeted."[72] If personal data processing for direct marketing can be based on the legitimate interests provision (article 7(f) of the Data Protection Directive), the directive grants data subjects the right to object to direct marketing – an opt-out regime.[73] This right is an unconditional right to object.[74] In sum: if, in rare circumstances, a company could rely on the legitimate interests provision for behavioural targeting, the data subject would have the right to stop the data processing: to opt out.

---

[69] See J Koëter, 'Behavioral Targeting en Privacy: een Juridische Verkenning van Internetgedragsmarketing' [Behavioural Targeting and Privacy: a Legal Exploration of Behavioural Internet Marketing] (2009) Tijdschrift voor Internetrecht 2009-4, p. 109-111.
[70] Information Commissioner, 'Personal information online. Code of practice' 2010. <https://ico.org.uk/media/for-organisations/documents/1591/personal_information_online_cop.pdf> accessed 13 May 2015.
[71] Article 29 Working Party, 'Opinion 03/2013 on purpose limitation' (WP 203), 2 April 2013, p. 46.
[72] Capitalisation adapted. The Working Party approved the code in Article 29 Working Party, 'Opinion 4/2010 on the European Code of Conduct of FEDMA for the Use of Personal Data in Direct Marketing' (WP 174), 13 July 2010.

The European Commission proposal for a Data Protection Regulation copies the legitimate interests provision without major changes.[75] In March 2014, the European Parliament adopted a compromise text, prepared by the Parliament's LIBE Committee. The LIBE Compromise allows companies, under certain conditions, to rely on the legitimate interests provision for behavioural targeting with pseudonymous data (data about individuals without a name attached).[76] The Working Party warns that the LIBE Compromise could be misunderstood as allowing companies to base most behavioural targeting practices on the legitimate interests provision, as long as companies use pseudonymous data.[77] The debate on behavioural targeting and necessity for the controller's legitimate interests as a legal basis has not been settled in Brussels.

In conclusion, under current law, personal data processing for behavioural targeting, in particular if it involves tracking people over multiple websites or Internet services, generally cannot be based on necessity for the legitimate interests of the controller.

---

[73] See article 14(b) and recital 30 of the Data Protection Directive.
[74] See Article 29 Working Party, 'Opinion 03/2013 on purpose limitation' (WP 203), 2 April 2013, p. 35; D Korff, *Data Protection Laws in the European Union* (Federation of European Direct Marketing and Direct Marketing Association 2005), p 100.
[75] But see Purtova, who argues that the proposal tilts the balance in favour of data controllers in the new version of the legitimate interests provision (N Purtova, 'Default entitlements in personal data in the proposed Regulation: Informational self-determination off the table… and back on again?' (2014) 30(1) Computer Law & Security Review 6).
[76] See article 2(a), article 6(f), and recitals 38 and 58a of the LIBE Compromise, proposal for a Data Protection Regulation (2013). The LIBE Compromise also requires a "highly visible" opt-out possibility (article 20(1); see also article 19(2)). This paper refers to the inofficial Consolidated Version after LIBE Committee Vote, provided by the Rapporteur, Regulation Of The European Parliament And Of The Council On The Protection Of Individuals With Regard To The Processing Of Personal Data And On The Free Movement Of Such Data (general Data Protection Regulation), 22 October 2013, <www.janalbrecht.eu/fileadmin/material/Dokumente/DPR-Regulation-inofficial-consolidated-LIBE.pdf> accessed 13 May 2015.
[77] Article 29 Working Party 2013, 'Draft Working Party comments to the vote of 21 October 2013 by the European Parliament's LIBE Committee' (12 November 2013) <http://ec.europa.eu/justice/data-protection/article-29/documentation/other-document/files/2013/20131211_annex_letter_to_greek_presidency_wp29_comments_outcome_vote_libe_final_en.pdf> accessed 13 May 2015.

# DATA SUBJECT'S CONSENT FOR PERSONAL DATA PROCESSING

If a firm cannot base personal data processing on the legitimate interests provision (article 7(f) of the Data Protection Directive) or another legal basis, only the data subject's consent can provide a legal basis for processing (article 7(a) of the Data Protection Directive). For valid consent, the Directive requires a (i) freely given, (ii) specific, (iii) informed (iv) indication of wishes, by which the data subject signifies agreement to his or her personal data being processed.[78] People can always withdraw their consent.[79]

If there is no indication of wishes there cannot be consent; so there is no need to check the other requirements for consent. As in general contract law, in data protection law an indication of wishes can be given in any form, and also implicitly.[80] But without special circumstances, mere inactivity is not an indication of wishes. Article 7(a) of the Data Protection Directive emphasises that a clear indication of wishes is required for valid consent: the provision requires "unambiguous" consent.

The European Court of Justice confirms that valid consent requires an indication of wishes. For instance, a data controller cannot obtain consent by merely informing people about processing.[81] "Acknowledging prior notice", says the Advocate General, "is not the same as giving 'unambiguous' consent (…). Nor can it properly be described as a 'freely given specific indication' of the [data subjects'] wishes in accordance with the definition of the data subject's consent in article 2(h)."[82] In another case, the Court suggests that "consent" in data protection law requires "express" consent.[83]

---

[78] Article 2(h) of the Data Protection Directive.
[79] European Commission, Amended proposal for a Council Directive on the Protection of Individuals with regard to the Processing of Personal Data and on the Free Movement of Such Data, COM (92) 422 final – SYN 287, 15 October 1992 [1992] OJ C311/30 (27.11.1992), p. 2. See also E Kosta, *Consent in European Data Protection Law* (PhD thesis University of Leuven) (Martinus Nijhoff Publishers 2013) p. 251, with further references.
[80] Idem, p. 368; C Kuner, *European Data Protection Law: Corporate Regulation and Compliance* (Oxford University Press 2007), p. 68; Article 29 Working Party, 'Opinion 15/2011 on the definition of consent' (WP 187) 13 July 2011, p. 11.
[81] CJEU, Case C-92/09 and C-93/09 Volker und Markus Schecke and Eifert [2010] ECLI:EU:C:2010:662, par. 63.

In case law outside the data protection field, the Court affirms that consent cannot be inferred from inactivity. In a case where the European Commission did not initiate an infringement procedure, this inactivity "cannot be interpreted as the Commission's tacit consent."[84] In two trademark cases, "implied consent (…) cannot be inferred from (…) mere silence",[85] and "'consent' (…) must be so expressed that an intention to renounce a right is unequivocally demonstrated."[86]

In the UK, regulators and commentators seem to be more inclined to accept a system that allows people to object – an opt-out system – as a way of obtaining "implied" consent.[87] Viewing an opt-out system as sufficient to obtain consent has been met with criticism in literature. For example, Kosta says "there is no such thing as 'opt-out consent'[88] (…) An 'opt-out' regime refers to the right of a data subject to object to the processing of his personal data and does not constitute consent."[89] Similarly, the Working Party says consent requires affirmative action.[90]

The difference between direct marketing that is based on the controller's legitimate interests under article 7(f) of the Data Protection Directive (on an opt-out basis) and direct marketing that is based on the data subject's consent under article 7(a) of that same directive (opt-in) isn't merely theoretical. The legitimate interests provision sometimes allows companies to process personal data for direct marketing on an opt-out basis – but in such cases the company must balance its interests against those of the data subject. By relying on fictitious opt-out consent, companies could try to escape that responsibility.[91]

---

[82] Opinion AG (ECLI:EU:C:2010:353) for CJEU, Case C-92/09 and C-93/09 Volker und Markus Schecke and Eifert [2010] ECLI:EU:C:2010:662, par. 79.
[83] The CJEU suggests that "consent" in Regulation (EC) No 45/2001 requires "express" consent (CJEU, Case C-28/08 and T-194/04 Bavarian Lager [2010] ECLI:EU:C:2010:378, par. 77. Article 2(h) of Regulation (EC) No 45/2001 uses the same consent definition as the Data Protection Directive. In another case, the CJEU reads "an opportunity to determine" as requiring "prior", "free, specific and informed consent" (CJEU, Case C-543/09 Deutsche Telekom [2011] ECLI:EU:C:2011:279, par. 55-58).
[84] CJEU, Case C-577/08 Brouwer [2010] ECLI:EU:C:2010:449, par. 39.
[85] ECJ, Case C-414/99 to C-416/99 Zino Davidoff [2001] ECLI:EU:C:2001:617, par. 55.
[86] CJEU, Case C-482/09 Budějovický Budvar [2011] ECLI:EU:C:2011:605, par. 42-44.
[87] See: D De Lima and A Legge, 'The European Union's approach to online behavioural advertising: Protecting individuals or restricting business?' (2014) 30(1) Computer Law & Security Review 67.
[88] P Traung, 'The Proposed New EU General Data Protection Regulation: Further Opportunities' (2012)(2) Computer Law Review international 33; E Kosta, *Consent in European Data Protection Law (PhD thesis University of Leuven)* (Martinus Nijhoff Publishers 2013), p. 202.
[89] Idem, p. 387.
[90] Article 29 Working Party, 'Working Document 02/2013 providing guidance on obtaining consent for cookies' (WP 208) 2 October 2013, p. 3.

A number of larger behavioural targeting companies, cooperating in the Interactive Advertising Bureau, offer people the chance to opt out of targeted advertising on a centralised website: youronlinechoices.com. But the companies merely promise to stop showing targeted advertising after people opt out: "Declining behavioural advertising only means that you will not receive more display advertising customised in this way."[92] Hence, companies may continue to collect data about people who opted out. The website thus offers the equivalent of Do Not Target, rather than Do Not Track. But even if companies stopped tracking people after an opt-out, it would be hard to see how the opt-out system could meet data protection law's requirements for consent.[93] In sum, valid consent requires an indication of wishes by the data subject.

The Data Protection Directive also requires consent to be "specific" and "informed." For instance, consent to use personal data "for commercial purposes" would not be specific.[94] Consent can only be informed if a consent request clearly explains how the company wants to use the data. In a case on working hours (not regarding data protection law), the European Court of Justice required "full knowledge of all the facts" for consent.[95]

Obtaining data subject consent must be distinguished from data protection law's requirements regarding information to be given to the data subject. Those requirements

---

[91] However, the legal basis consent does not legitimise excessive data processing (article 6(c) of the Data Protection Directive).
[92] Interactive Advertising Bureau, 'Your Online Choices. A Guide to Online Behavioural Advertising. FAQ 22' <www.youronlinechoices.com/ma/faqs#22> accessed 14 May 2015.
[93] Article 29 Working Party, 'Opinion 16/2011 on EASA/IAB Best Practice Recommendation on Online Behavioural Advertising' (WP 188) 8 December 2011, p. 6.
[94] European Commission, Amended proposal for a Council Directive on the Protection of Individuals with regard to the Processing of Personal Data and on the Free Movement of Such Data, COM (92) 422 final – SYN 287, 15 October 1992 [1992] OJ C311/30 (27.11.1992), p. 15; See also Article 29 Working Party, 'Opinion 15/2011 on the definition of consent' (WP 187) 13 July 2011, p. 17.
[95] ECJ, Case C-397/01 and C-403/01 Pfeiffer and others [2004] ECLI:EU:C:2004:584, dictum (2) and par. 82.

also apply when data controllers invoke another legal basis than consent.[96] Hence, even when a data controller does not seek the data subject's consent, it must provide certain information to data subjects, for instance with a privacy statement. It is not possible to obtain consent by silently changing a privacy statement or terms and conditions. If a data subject does not know about the new privacy statement, there cannot be an indication of wishes. "Consent must be specific", says the Working Party. "Rather than inserting the information in the general conditions of the contract, this calls for the use of specific consent clauses, separated from the general terms and conditions"[97]

Consent must be "freely given", so consent given under too much pressure is not valid. For instance, the European Court of Justice says that if people can only obtain a new passport if they give their fingerprints, those people cannot be deemed to have "freely given" consent, because a real choice is lacking. After all, people need a passport.[98] However, current data protection law does not explicitly prohibit controllers from offering take-it-or-leave-it choices. Hence, in principle website publishers are allowed to install tracking walls that deny entry to visitors that do not consent to being tracked.[99] Scholars are divided on how much pressure makes consent involuntary.[100] The Working Party does not like tracking walls, but does not say that current law prohibits them.[101]

In sum, the data subject's unambiguous consent (article 7(a) of the Data Protection Directive) is generally the required legal basis for personal data processing for behavioural targeting. The next section discusses a different consent requirement: the e-Privacy Directive's consent requirement for storing or accessing information on a user's or subscriber's device.

---

[96] See article 10 and 11 of the Data Protection Directive. Article 11 requires information "where the data have not been obtained from the data subject." In such cases there's no consent.
[97] Article 29 Working Party, 'Opinion 15/2011 on the definition of consent' (WP 187) 13 July 2011, p. 33-35.
[98] CJEU, Case C-291/12 Schwartz v Stadt Bochum [2013] ECLI:EU:C:2013:670, par. 32.
[99] See N Helberger, 'Freedom of Expression and the Dutch Cookie-Wall' (2013) March Institute for Information Law <www.ivir.nl/publicaties/download/1091> accessed 13 May 2015; FJ Zuiderveen Borgesius, 'Informed Consent: We Can Do Better to Defend Privacy' (2015) IEEE Security and Privacy (In Our Orbit) (2015) 13-2 103.
[100] See: A Roosendaal, *Digital personae and profiles in law: Protecting individuals' rights in online contexts (PhD thesis University of Tilburg, Academic version)* (2013) <http://ssrn.com/abstract=2313576> accessed 13 May 2015; E Kosta, 'Peeking into the cookie jar: the European approach towards the regulation of cookies' (2013) International Journal of Law and Information Technology 1, p. 17.
[101] Article 29 Working Party, 'Working Document 02/2013 providing guidance on obtaining consent for cookies' (WP 208) 2 October 2013, p. 5.

# E-PRIVACY DIRECTIVE: CONSENT FOR STORING AND ACCESSING INFORMATION ON A DEVICE

European legal discussions on behavioural targeting tend to focus on the e-Privacy Directive's consent requirement for tracking cookies and similar technologies, rather than on the Data Protection Directive. In short, article 5(3) of the e-Privacy Directive requires anyone who stores or accesses information on the device of a user or subscriber to obtain the consent of that user or subscriber, unless an exception applies.

Article 5(3) has several rationales. First, article 5(3) is part of article 5, which concerns "confidentiality of the communications". Article 5(1) protects the confidentiality of communications and the related traffic data, and applies, for instance, to phone calls and to email traffic. It makes sense that the law also protects a message after somebody downloaded and stored the message. Indeed, article 5(3) extends the right to communications confidentiality, and protects the contents of a device of a user or subscriber.[102] The e-Privacy Directive's preamble says that a user's device and its contents is part of the user's private sphere. "Terminal equipment of users of electronic communications networks and any information stored on such equipment are part of the private sphere of the users requiring protection under the European Convention for the Protection of Human Rights and Fundamental Freedoms."[103] Therefore, such devices and their contents, such as saved messages and address books, may only be accessed with the user's or subscriber's consent.[104]

---

[102] See W Steenbruggen W, *Publieke dimensies van privé-communicatie: een onderzoek naar de verantwoordelijkheid van de overheid bij de bescherming van vertrouwelijke communicatie in het digitale tijdperk (Public dimensions of private communication: an investigation into the responsibility of the government in the protection of confidential communications in the digital age) (PhD thesis University of Amsterdam)* (Academic version 2009), p. 186.
[103] Recital 24 of the e-Privacy Directive.

The preamble suggests that another rationale for article 5(3) is protecting users and subscribers against secretly installing information on their devices. The provision aims, for instance, to protect people against adware or spyware.[105] Article 5(3) also aims to protect people against surreptitious tracking with cookies and similar files.[106] While article 5(3) applies to storing or accessing any information on people's devices, for ease of reading this paper also speaks of consent for cookies.

Early proposals for the 2002 version of the e-Privacy Directive required companies to ask for consent before they placed certain kinds of cookies. After lobbying by the marketing industry, the final version used ambiguous wording about a "right to refuse." The 2002 version of article 5(3) is usually interpreted as granting a right to object to cookies (an opt-out system).[107]

Since 2009, article 5(3) of the e-Privacy Directive requires any party that stores or accesses information on a user's device to obtain the user's informed consent.[108] For the definition of consent, the e-Privacy Directive refers to the Data Protection Directive.[109] Article 5(3) provides for exceptions to the consent requirement, for example for cookies that are necessary for transmitting communication or for a service requested by the user. Hence, no prior consent is needed for cookies that are used for log-in procedures or for digital shopping carts.

---

[104] The Working Party confirms that the provision applies, for instance, to apps that access information on a user's smartphone, such as location data or a user's contact list (Article 29 Working Party 2013, WP 202 – 'Opinion 02/2013 on apps on smart devices' (WP 202) 27 February 2013, p. 10).
[105] Recital 24 of the e-Privacy Directive.
[106] See e.g. recital 24 and 25 of the e-Privacy Directive, and recital 65 and 66 of Directive 2009/136.
[107] S Kierkegaard, 'How the cookies (almost) crumbled: privacy & lobbyism' (2005) 21(4) Computer Law & Security Review 310. Some authors read the 2002 version as requiring (prior) consent (see e.g. Traung P, 'EU Law on Spyware, Web Bugs, Cookies, etc. Revisited: Article 5 of the Directive on Privacy and Electronic Communications' (2010) 31 Business Law Review 216, with further references).
[108] The e-Privacy Directive 2002/58 was updated by Directive 2009/136. This paper refers to the consolidated version from 2009.
[109] Article 2(f) and recital 17 of the e-Privacy Directive.

A recital of the 2009 directive that amended the e-Privacy Directive has caused much discussion: "in accordance with the relevant provisions of [the Data Protection Directive], the user's consent to processing may be expressed by using the appropriate settings of a browser or other application."[110] Some marketers suggest that people who don't block tracking cookies in their browser give implied consent to behavioural targeting. For instance, the Interactive Advertising Bureau UK says "<u>default</u> web browser settings can amount to 'consent'."[111] But this interpretation of the law is not convincing. As the Working party notes, people do not indicate that they allow tracking cookies on their devices, merely because they leave their browsers' default settings untouched.[112] In addition, if default browser settings could indicate consent, this would imply that people consented to spyware and viruses if they did not block such files in their browsers. It is implausible that the European lawmaker had such an interpretation in mind.[113]

The 2009 version of article 5(3) should have been implemented in national legislation in May 2011, but many member states missed this deadline.[114] At the time of writing, enforcement of the consent requirement for tracking cookies is in its infancy, among other reasons because the national laws implementing article 5(3) are rather new.[115]

---

[110] Directive 2009/136, recital 66.
[111] Interactive Advertising Bureau United Kingdom, 'Department for Business, Innovation & Skills consultation on implementing the revised EU electronic communications framework, IAB UK Response' (1 December 2012) <www.iabuk.net/sites/default/files/IABUKresponsetoBISconsultationonimplementingtherevisedEUElectronicCommunicationsFramework_7427_0.pdf> accessed 13 May 2015, p. 2
[112] See e.g. Article 29 Working Party, 'Opinion 15/2011 on the definition of consent' (WP 187) 13 July 2011, p. 32; p. 35.
[113] See N Van Eijk et al, 'A bite too big: Dilemma's bij de implementatie van de Cookiewet in Nederland' [A bite too big, dilemmas with the implementation of the cookie law in the Netherlands] (2011) Report nr. 35473 TNO/IViR <http://ivir.nl/publicaties/download/228 > accessed 14 May 2015, p. 63.
[114] Article 4(1) of Directive 2009/136. According to the Working Party, all member states had implemented the amended e-Privacy Directive on 1 January 2013 (Article 29 Working Party, 'Working Document 02/2013 providing guidance on obtaining consent for cookies' (WP 208) 2 October 2013, p. 2). It's not unusual that member states implement directives late.
[115] Regulators have taken some action regarding the national implementation of article 5(3). For example, the Spanish Data Protection Authority issued a fine for non-compliance in January 2014 (Agencia Española de Protección de Datos, 'decision regarding Navas Joyeros Importadores, S.l. Y Privilegia Luxury Experience, S.l.', 15 January 2014. PS/00321/2013 <www.agpd.es/portalwebAGPD/resoluciones/procedimientos_sancionadores/ps_2014/common/pdfs/PS-00321-2013_Resolucion-de-fecha-14-01-2014_Art-ii-culo-5.1-LOPD-22.2-LSSI.pdf> accessed 13 May 2015). The Dutch Data Protection Authority has concluded in several investigations that article 5(3)

It is unclear how national authorities will apply the national implementation of article 5(3). The approaches seem to vary. For instance, the UK appears to accept opt-out systems to obtain consent for cookies,[116] while the Netherlands requires an indication of wishes for valid consent.[117] The Working Party consistently says that "an active indication of the user's wishes" is required for consent to cookies.[118] A report on the national implementation of the e-Privacy Directive has been completed for the European Commission, but has not been published yet.[119]

# THE E-PRIVACY DIRECTIVE'S COOKIE PROVISION DOES NOT PROVIDE A LEGAL BASIS FOR PERSONAL DATA PROCESSING

How should the relation between article 5(3) of the e-Privacy Directive and the legal basis requirement of article 7 of the Data Protection Directive be interpreted? The answer mainly depends on whether one sees article 5(3) of the e-Privacy Directive as

---

was breached (see Dutch Data Protection Authority - Annual Report 2014 (English summary) <https://cbpweb.nl/sites/default/files/atoms/files/annual_report_2014.pdf> accessed 14 May 2015.
[116] For at least one year the English Information Commissioner's Office (ICO), the regulator that oversees compliance with the e-Privacy Directive, dropped cookies through its website as soon as a visitor arrived, and explained in a banner that it had done so. The ICO appeared to suggest that explaining how a user can delete cookies is enough to obtain "implied" consent ((Information Commissioner's Office, 'Changes to cookies on our website' (31 January 2013) <https://ico.org.uk/about-the-ico/news-and-events/current-topics/changes-to-cookies-on-our-website/> accessed 13 May 2015).
[117] See R Leenes R and E Kosta, 'Taming the cookie monster with Dutch law – A tale of regulatory failure' (2015) Computer Law & Security Review, 31.2.
[118] Article 29 Working Party, 'Working Document 02/2013 providing guidance on obtaining consent for cookies' (WP 208) 2 October 2013, p. 3.
[119] 'E-Privacy Directive: assessment of transposition, effectiveness and compatibility with proposed Data Protection Regulation' <www.sparklegalnetwork.eu/spark-and-time-lex-study-completed> accessed 13 May 2015. Several law offices have also compiled overviews of the implementation of article 5(3). See e.g. Bird & Bird <www.twobirds.com/~/media/PDFs/Expertise/Data%20Protection/Twobirds%20Cookie%20Directive%20Tracker%202014.PDF>; DLA Piper <www.dlapiper.com/~/media/Files/Insights/Publications/2014/09/EU_Cookies_Update_September_2014.pdf>; Field Fisher <www.fieldfisher.com/media/2927368/EU-Cookie-Consent-Tracking-Table-Fieldfisher-21-April-2015.pdf> accessed 13 May 2015.

*complementing*, or as *particularising* article 7 of the Data Protection Directive. As several authors note, the e-Privacy Directive could have been clearer on this point.[120]

I argue that article 5(3) of the e-Privacy Directive does not provide a legal basis for personal data processing, as article 5(3) concerns a different topic than personal data. Article 5(3) of the e-Privacy Directive *complements* article 7 of the Data Protection Directive, because the two provisions concern different subject matter. Therefore, the two provisions both apply if a company uses a tracking cookie to process personal data for behavioural targeting.

The e-Privacy Directive says in article 1(2) that its provisions "particularise and complement" the Data Protection Directive. Several provisions of the e-Privacy concern topics outside the scope of the Data Protection Directive. For instance, the e-Privacy Directive not only protects the interests of natural persons, but also of legal persons.[121] Protecting legal persons goes beyond the scope of the Data Protection Directive, because the latter does not protect data about legal persons.[122] Provisions in the e-Privacy Directive that deal with matters outside the scope of the Data Protection Directive *complement* the Data Protection Directive. A provision that complements the Data Protection Directive adds an extra rule – the provision does not particularise, or specify, a rule of the Data Protection Directive.

One interpretation – not persuasive in my opinion – is that article 5(3) of the e-Privacy Directive *particularises* (rather than complements) article 7 of the Data Protection Directive. The result of that interpretation would arguably be as follows: a company that obtains a subscriber's or user's consent for storing a tracking cookie (article 5(3) of the e-Privacy Directive) automatically obtains a data subject's "unambiguous consent" as a legal basis for subsequent personal data processing (article 7(a) of the Data Protection Directive).[123]

---

[120] See e.g. P Rosier, 'Comments on the Data Protection Directive' in Büllesbach A et al (eds), *Concise European IT Law (second edition)* (Kluwer Law International 2010), p. 176; F Debusseré, 'The EU E-Privacy Directive: A Monstrous Attempt to Starve the Cookie Monster' (2005) 13 International Journal of Law and Information Technology 70; W Kotschy, 'The proposal for a new General Data Protection Regulation—problems solved?' (2014) International Data Privacy Law 4(4) 274; V Papakonstantinou and P De Hert, 'The Amended EU Law on ePrivacy and Electronic Communications after Its 2011 Implementation; New Rules on Data Protection, Spam, Data Breaches and Protection of Intellectual Property Rights' (2011) 29 John Marshall Journal of Computer & Information Law, p. 42.
[121] Article 1(2) of the e-Privacy Directive.
[122] Article 2(a) and Article 3(1) of the Data Protection Directive. In some cases general data protection law can apply to data about legal persons. See B Van der Sloot, 'Do privacy and data protection rules apply to legal persons and should they? A proposal for a two-tiered system' (2015) 31(1) Computer Law & Security Review 26; LA Bygrave, *Data Protection Law: Approaching its Rationale, Logic and Limits (PhD thesis University of Oslo)* (Information Law Series, Kluwer Law International 2002), part III.

An argument for that interpretation is that the e-Privacy Directive is often called a *lex specialis* of the Data Protection Directive.[124] For example, in 2010 the Working Party still suggested that the e-Privacy Directive was a *lex specialis* of the Data Protection Directive.[125] (We will see below that the Working Party has changed its view – or perhaps the Working Party never meant to say that article 5(3) is a *lex specialis*.) If the *whole* e-Privacy Directive were a *lex specialis* of the Data Protection Directive, article 5(3) of the e-Privacy Directive could arguably be seen as a *lex specialis* of article 7 of the Data Protection Directive.

I argue that article 5(3) of the e-Privacy Directive does not *particularise* article 7 of the Data Protection Directive, because the two provisions have a different scope, a different focus, and a different goal. In other words, article 5(3) of the e-Privacy Directive should not be seen as a *lex specialis* of article 7 of the Data Protection Directive. The phrase *lex specialis derogat legi generali* implies: "whenever two or more norms deal with the same subject matter, priority should be given to the norm that is more specific."[126] As article 5(3) of the e-Privacy Directive does not deal with the same subject matter as article 7 of the Data Protection Directive, article 5(3) of the e-Privacy Directive is not a *lex specialis* of article 7 of the Data Protection Directive. More generally, Kotschy says that the fact that the e-Privacy Directive "complements and particularises" the Data

---

[123] See: F Debusseré, 'The EU E-Privacy Directive: A Monstrous Attempt to Starve the Cookie Monster' (2005) 13 International Journal of Law and Information Technology 70, p. 95.
[124] See for instance: European Commission, Communication of 6 May 2015 on A Digital Single Market for Europe, COM (2015) 192 final, p. 47.
[125] Article 29 Working Party, 'Opinion 2/2010 on online behavioural advertising' (WP 171), 22 June 2010, p. 9-10.
[126] M. Koskenniemi, 'The function and scope of the lex specialis rule and the question of self-contained regimes', in Koskenniemi et al, *Report of the Study Group of the International Law Commission*, A/CN.4/L.702, 18 July 2006 < http://www.wsi.uni-kiel.de/de/lehre/vorlesungen/archiv/ss-2014/arnauld/voelkerrecht-ii/materialien/ilc-fragmentation-a-cn.4-l.702-2006> accessed 13 May 2015, p. 8. See on the interpretation of *lex specialis* also Advocate General Sharpston, Case C-355/12 Nintendo [2013] ECLI:EU:C:2013:581.

Protection Directive does not imply a *lex specialis/lex generalis* relation between the two directives.[127] Here, I focus only on article 5(3) of the e-Privacy Directive.

There are several arguments for viewing article 5(3) of the e-Privacy Directive as *complementing* (rather than particularising) the Data Protection Directive. First, the text of article 5(3) shows that the provision applies regardless of whether personal data are processed; article 5(3) applies to "storing of information, or the gaining of access to information already stored". Article 5(3) does not use the phrase "personal data", and the recitals confirm that article 5(3) also applies when no personal data are processed.[128] For instance, article 5(3) applies to installing viruses and spyware – that has little to do with personal data.[129] The relevant recitals mention privacy, but not personal data processing.[130]

At first glance, article 3 of the e-Privacy Directive might give the impression that article 5(3) only applies when personal data are stored or accessed on the device of a subscriber or user. Article 3 appears to suggest that the e-Privacy Directive only applies when personal data are being processed: "[t]his Directive shall apply to the *processing of personal data* in connection with the provision of publicly available electronic communications services (…)."[131]

However, the e-Privacy Directive clearly applies in some situations that do not involve personal data. For instance, the e-Privacy Directive states it also protects the interests of subscribers who are legal persons (article 1(2)).[132] Apparently, article 3 does not imply that the e-Privacy Directive only applies if personal data are processed. In sum, article 5(3) also applies when information other than personal data is stored or accessed on the device of a subscriber or user.

---

[127] W Kotschy, 'The proposal for a new General Data Protection Regulation—problems solved?' (2014) International Data Privacy Law 4(4) 274.
[128] See recital 25 and 25 of the 2002 e-Privacy Directive, and recital 65 and 66 of Directive 2009/136 (that amended the e-Privacy Directive).
[129] Recital 24 of the e-Privacy Directive.
[130] See recital 25 and 25 of the 2002 e-Privacy Directive, and recital 65 and 66 of Directive 2009/136 (that amended the e-Privacy Directive).
[131] Emphasis added.
[132] See also recital 12 of the e-Privacy Directive. A "subscriber" is defined in article 2(k) of the Framework Directive 2002/21.

A second reason to see article 5(3) of the e-Privacy Directive as complementing (not particularising) the Data Protection Directive relates to the goal of article 5(3), and of article 5 more generally. Article 5(3) of the e-Privacy Directive has a different goal than article 7 of the Data Protection Directive. As said, article 5 concerns the confidentiality of communications, and article 5(3) of the e-Privacy Directive aims to protect natural and legal persons against surreptitious tracking, against spyware or viruses, and against unauthorized access of their devices.[133] In contrast, article 7 of the Data Protection Directive only concerns the legal basis for the processing of personal data.

A third argument to interpret article 5(3) of the e-Privacy Directive as complementing the Data Protection Directive follows from recital 10 of the e-Privacy Directive. That recital confirms that some situations in the electronic communications sector are not regulated by the e-Privacy Directive – if personal data are processed in such situations, the Data Protection Directive applies. Recital 10 says: "[i]n the electronic communications sector, [the Data Protection] Directive 95/46/EC applies in particular to all matters concerning protection of fundamental rights and freedoms, which are not specifically covered by the provisions of this [e-Privacy] Directive, including the obligations on the controller and the rights of individuals." Under the Data Protection Directive, one obligation of a data controller is having a legal basis for processing (article 7).[134]

In a 2013 opinion, the Working Party suggests that a company that obtains consent under article 5(3) of the e-Privacy Directive does not automatically obtain a legal basis for subsequent personal data processing (article 7(a) of the Data Protection Directive): "It is important to note the distinction between the consent required to place any information on and read information from the device, and the consent necessary to have a legal ground for the processing of different types of personal data."[135] Hence, the Working Party confirms that 5(3) of the e-Privacy Directive and article 7 of the Data Protection Directive deal with separate topics.

---

[133] Recital 24 and 25 of the e-Privacy Directive.
[134] See article 7 of the Data Protection Directive.

*Implications*

The foregoing seems to allow only one conclusion: article 5(3) of the e-Privacy Directive complements – not particularises – article 7 of the Data Protection Directive. Article 5(3) of the e-Privacy Directive thus does not provide a legal basis for personal data processing.

If correct, that conclusion has several implications. For instance, if a company obtains consent from a subscriber (that may be a legal person) or user for storing or accessing a tracking cookie on the device of the subscriber or user, the company still needs a legal basis if it processes personal data for behavioural targeting. As discussed above, in most cases only the data subject's "unambiguous consent" (article 7 of the Data Protection Directive) can provide a legal basis for the processing of personal data for behavioural targeting.

In such cases, two separate provisions require the company to obtain consent, for two different activities. First, under article 5(3) of the e-Privacy Directive, the company must obtain consent of the user or subscriber for storing or reading a tracking cookie. Second, under article 7(a) of the Data Protection Directive, the company (as a data controller) must obtain the relevant data subject's unambiguous consent for the processing of personal data for behavioural targeting. At first glance, this appears to imply that the company must ask for consent twice.

But it makes sense to combine the two consent requests. The Working Party said in 2013: "[t]hough both consent requirements are simultaneously applicable (…) the two types of consent can be merged in practice (…)."[136] Outside the field of data protection law, two indications of wishes are sometimes combined as well. For instance, somebody can buy a phone and insurance for that phone – by signing one contract.

---

[135] Article 29 Working Party, 'Opinion 02/2013 on apps on smart devices' (WP 202) 27 February 2013, p. 14.
[136] Idem, p. 14.

Hence: even if companies could use an opt-out system to obtain "implied" consent (of the user or subscriber) for using a tracking cookie, they would generally have to obtain unambiguous consent (of the data subject) if they process personal data for behavioural targeting. While it is already difficult to see how an opt-out system could be used to obtain consent for cookies, it would be even more difficult to see how an opt-out system could be used to obtain a data subject's "unambiguous" consent. After all, an indication of wishes is required for valid consent – and mere silence of the data subject is not an indication of wishes.

If – in rare cases – a company could base the processing of personal data for behavioural targeting on the legitimate interests provision (article 7(f) of the Data Protection Directive), the company would still have to obtain consent for using the tracking cookie (article 5(3) of the e-Privacy Directive). From the company's perspective, it is thus hardly relevant which legal basis it can rely upon for personal data processing for behavioural targeting, as operating the tracking cookie requires consent.[137]

The fact that article 5(3) of the e-Privacy Directive does not provide a legal basis for personal data processing has a further implication. Even in the hypothetical situation that article 5(3) of the e-Privacy Directive were abolished, companies would generally have to obtain unambiguous consent of Internet users before tracking them for behavioural targeting. However, that conclusion only applies when behavioural targeting entails personal data processing. As noted, according to the Working Party, behavioural targeting usually entails the processing of personal data.[138]

---

[137] See Article 29 Working Party, 'Opinion 06/2014 on the notion of legitimate interests of the data controller under article 7 of Directive 95/46/EC' (WP 217) 9 April 2014, p. 46.
[138] See e.g. Article 29 Working Party, 'Opinion 2/2010 on online behavioural advertising' (WP 171), 22 June 2010, p. 9.

## CONCLUSION

In conclusion, the cookie consent requirement of the e-Privacy Directive does not provide a legal basis for the processing of personal data. As far as behavioural targeting entails personal data processing, the data controller that uses behavioural targeting, such as an advertising network, needs a legal basis for the processing. The European Union Charter of Fundamental Rights only allows personal data processing if it can be based on the data subject's consent, or on another legal basis. For the private sector the most relevant legal bases are: necessity for performance of a contract, necessity for the controller's legitimate interests, and the data subject's unambiguous consent. Usually companies cannot base the processing of personal data for behavioural targeting on the legal basis necessity for performance of a contract, or on the legal basis necessity for the controller's legitimate interests. Therefore, the processing of personal data for behavioural targeting almost always requires the data subject's unambiguous consent.

∗ ∗ ∗